\documentclass[10pt,twocolumn,journal]{IEEEtran}

\usepackage{amsmath, mathrsfs,balance}
\usepackage{bm}

\usepackage{graphicx}
\usepackage{color}
\usepackage{tikz}
\usepackage{pgfplots}
\pgfplotsset{compat=newest}
\usetikzlibrary{patterns}
\usetikzlibrary{positioning}
\usetikzlibrary{datavisualization}
\usetikzlibrary{datavisualization.formats.functions}
\usetikzlibrary{backgrounds}
\usetikzlibrary{shapes,snakes}

\makeatletter
\def\maketag@@@#1{\hbox{\m@th\normalfont\normalsize#1}}
\makeatother

\usepackage[none]{hyphenat}

\newcommand{\subparagraph}{}
\usepackage[compact]{titlesec}
\titlespacing*{\section}{2pt}{1\baselineskip}{0.9\baselineskip}

\allowdisplaybreaks

\usepackage{amsfonts}
\usepackage{amssymb}
\usepackage{color}
\usepackage{multirow}
\usepackage{graphicx}

\usepackage{caption}
\usepackage{subcaption}
\usepackage[numbers,sort&compress]{natbib}
\usepackage{balance}

\usepackage{mathbbol}

\def\mindex#1{\index{#1}}

\def\sq{\hbox{\rlap{$\sqcap$}$\sqcup$}}
\def\qed{\ifmmode\sq\else{\unskip\nobreak\hfil
\penalty50\hskip1em\null\nobreak\hfil\sq
\parfillskip=0pt\finalhyphendemerits=0\endgraf}\fi\medskip}

\long\def\defbox#1{\framebox[.9\hsize][c]{\parbox{.85\hsize}{%
\parindent=0pt
\baselineskip=12pt plus .1pt      
\parskip=6pt plus 1.5pt minus 1pt 
 #1}}}

\long\def\beginbox#1\endbox{\subsection*{}%
\hbox{\hspace{.05\hsize}\defbox{\medskip#1\bigskip}}%
\subsection*{}}

\def\endbox{}

\newsavebox{\junk}
\savebox{\junk}[1.6mm]{\hbox{$|\!|\!|$}}

\def\argmin{\mathop{\rm arg\, min}}

\newcommand{\field}[1]{\mathbb{#1}}

\def\Re{\field{R}}

\def\intgr{\field{Z}}

\def\Co{\field{C}}

\def\bC{{\mathbb C}}

\def\bE{{\mathbb E}}

\def\bR{{\mathbb R}}

\def\bfA{{\bf A}}
\def\bfB{{\bf B}}

\def\bfF{{\bf F}}

\def\bfH{{\bf H}}

\def\bfa{{\bf a}}
\def\bfb{{\bf b}}
\def\bfc{{\bf c}}

\def\bfg{{\bf g}}
\def\bfh{{\bf h}}

\def\bfq{{\bf q}}
\def\bfr{{\bf r}}

\def\bfu{{\bf u}}
\def\bfv{{\bf v}}
\def\bfw{{\bf w}}
\def\bfx{{\bf x}}

\def\tto{{\mathtt o}}

\def\ttt{{\mathtt t}}

\def\sfH{{\sf H}}

\def\bfmath#1{{\mathchoice{\mbox{\boldmath$#1$}}%
{\mbox{\boldmath$#1$}}%
{\mbox{\boldmath$\scriptstyle#1$}}%
{\mbox{\boldmath$\scriptscriptstyle#1$}}}}

\def\bfmY{\bfmath{Y}}

\def\bfmhhaY{\bfmath{\hhaY}} 
\def\bfmhhaY{\hbox to 0pt{$\widehat{\bfmY}$\hss}\widehat{\phantom{\raise 1.25pt\hbox{$\bfmY$}}}}

\def\til={{\widetilde =}}

\def\clC{{\cal C}}

\def\clN{{\cal N}}

 \def\FRAC#1#2#3{\genfrac{}{}{}{#1}{#2}{#3}}

\def\ddtp{{\mathchoice{\FRAC{1}{d^{\hbox to 2pt{\rm\tiny +\hss}}}{dt}}%
{\FRAC{1}{d^{\hbox to 2pt{\rm\tiny +\hss}}}{dt}}%
{\FRAC{3}{d^{\hbox to 2pt{\rm\tiny +\hss}}}{dt}}%
{\FRAC{3}{d^{\hbox to 2pt{\rm\tiny +\hss}}}{dt}}}}

\def\average#1,#2,{{1\over #2} \sum_{#1}^{#2}}

\def\eye(#1){{\bf(#1)}\quad}

\def\eq#1/{(\ref{e:#1})}

\newcommand{\beqn}[1]{\notes{#1}%
\begin{eqnarray} \elabel{#1}}

\newcommand{\eeqn}{\end{eqnarray} }

\newcommand{\beq}[1]{\notes{#1}%
\begin{equation}\elabel{#1}}

\newcommand{\eeq}{\end{equation}}

\def\bdes{\begin{description}}
\def\edes{\end{description}}

\newcounter{rmnum}

\newcounter{anum}

{\end{list}}

\def\ass(#1:#2){(#1\ref{#1:#2})}

\def\ritem#1{
\item[{\sf \ass(\current_model:#1)}]
}

\newenvironment{recall-ass}[1]{%
\begin{description}
\def\current_model{#1}}{
\end{description}
}

\long\def\comment#1{}

\newcommand{\uv}{{\bf u}}

\newcommand{\vv}{{\bf v}}

\newcommand{\Fm}{{\bf F}}

\newcommand{\Gammam}{\boldsymbol{\Gamma}}

\renewcommand{\Re}{{\rm Re}}

\newcommand{\transp}{{\sf T}}
\renewcommand{\vec}{{\rm vec}}

\def\snrbef{{\mathsf{SNR}_\text{BBF}}}

\usepackage{tikz}
\usepackage{pgfplots}
\pgfplotsset{compat=newest}
\usetikzlibrary{patterns}
\usetikzlibrary{positioning}
\usetikzlibrary{datavisualization}
\usetikzlibrary{datavisualization.formats.functions}
\usetikzlibrary{backgrounds}
\usetikzlibrary{shapes,snakes}
\usepackage{textcomp}
\usetikzlibrary{automata}

\allowdisplaybreaks

\usepackage{algorithm}
\usepackage{algpseudocode}
\usepackage{pifont}
\usepackage{varwidth}

\def\one{{\bf 1}}

\usepackage{multicol,lipsum}

\def\herm{{\sfH}}

\def\snr{{\mathsf{SNR}}}

\def\ptot{{P_{\ttt \tto \ttt}}}

\def\cg{{\clC\clN}} 
\def\matlab{{MATLAB\textcopyright\,}}

\usepackage{color}
\newcommand{\figref}[1]{Fig.~\ref{#1}}
\begin{document}

\title{A Robust Time-Domain Beam Alignment Scheme for Multi-User Wideband mmWave Systems}
\author{Xiaoshen Song, Saeid Haghighatshoar,  \IEEEmembership{Member, IEEE,} Giuseppe Caire, \IEEEmembership{Fellow, IEEE} \\
Communications and Information Theory Group (CommIT), Technische Universit{\"a}t Berlin\\
xiaoshen.song@campus.tu-berlin.de, \{saeid.haghighatshoar, caire\}@tu-berlin.de
}
\maketitle\begin{abstract}
Millimeter wave (mmWave) communication with large array gains is a key ingredient of next generation (5G) wireless networks. Effective communication in mmWaves usually depends on the knowledge of the channel. We refer to the problem of finding a narrow beam pair at the transmitter and at the receiver, yielding high {\em Signal to Noise Ratio} (SNR) as {\em Beam Alignment} (BA). Prior BA schemes typically considered deterministic channels, where the instantaneous channel coefficients are assumed to stay constant for a long time. In this paper, in contrast, we propose a time-domain BA scheme for wideband mmWave systems, where the channel is characterized by multi-path components, different delays, {\em Angle-of-Arrivals}/{\em Angle-of-Departures} (AoAs/AoDs), and Doppler shifts. In our proposed scheme, the {\em Base Station} (BS) probes the channel in the downlink by some sequences with good autocorrelation property (e.g., \textit{Pseudo-Noise} (PN) sequences), letting each user estimate its best AoA-AoD that connects the user to the BS with two-sided high beamforming gain. We leverage the sparse nature of mmWaves in the AoA-AoD-time domain, and formulate the BA problem as a {\em Compressed Sensing} (CS) of a non-negative sparse vector. We use the recently developed {\em Non-Negative Least Squares} (NNLS) technique to efficiently find the strongest path connecting the BS and each user. Simulation results show that the proposed scheme outperforms its counterpart in terms of the training overhead and robustness to fast channel variations.
\end{abstract}	
\section{Introduction}\label{introduction}
The ever increasing demand for wireless and mobile data has placed a huge strain on existing WiFi and cellular networks \cite{GlobalData2016,Katabi2016}. Millimeter Wave (mmWave) frequency bands are considered as a promising solution by offering multi-GHz unlicensed bandwidth and providing Gigabits-per-second data rates \cite{OverviewHeath2016}. However, the high propagation loss and the unfavorable atmospheric absorption make the data transmission over long distances a serious challenge at mmWaves. To provide sufficient receive power in mmWave systems, large antenna arrays with high beamforming gains are required both at the transmitter and the receiver \cite{OverviewHeath2016}. The problem of finding a narrow beam pair at the transmitter and at the receiver, yielding a high  {\em Signal to Noise Ratio} (SNR), is referred to here as {\em Beam Alignment} (BA). This procedure is crucial for acquiring channel state information, i.e., the best {\em Angle-of-Arrival} (AoA) and {\em Angle-of-Departure} (AoD), which would benefit the subsequent data transmission \cite{SaeidBA2016,sxsBA2017}. BA is also known to be quite challenging due to the {\em Hybrid-Digital-Analog} (HDA) structure constraints, i.e. the number of available {\em Radio Frequency} (RF) chains is limited and direct channel estimation based on traditional all-digital solutions is infeasible. Moreover, the fact that, in mmWaves, the SNR before beamforming is very low and the dimensions of the channel matrices are very large further complicates the BA problem. 


Due to the sparse nature of mmWave channels \cite{OverviewHeath2016}, {\em Compressed Sensing} (CS) is considered as a powerful BA technique to reduce the number of training slots. Prior work based on CS techniques considered mainly deterministic channels \cite{OverviewHeath2016}, where the instantaneous channel coefficients were assumed to be constant for a long time \cite{AhmedFreqOMP2015,Gaozhen2016,LeeTwoStage2016}. In practice, however, mmWave channels would be fast time-varying due to the large carrier frequency and large Doppler spread \cite{WeilerMeasure2014}. In \cite{sxsBA2017}, an efficient BA scheme based on OFDM signaling was proposed. While this scheme is robust to fast time-varying mmWave channels, its performance may degrade when the channel is frequency selective and the subcarriers are not appropriately selected. More recently, \cite{AhmedTime2017,AlkhateebTimeDomain2017} proposed a time-domain wideband BA approach based on {\em Orthogonal Matching Persuit} (OMP) technique. It works for single-carrier modulation, but again, the underlying assumption that instantaneous channel coefficients remain invariant for the whole training stage is difficult to meet in mmWaves. We observe here that it is important to incorporate into the model both the frequency selectivity and the fast time variation properties of the mmWave channel, in order to develop a more practically relevant and robust BA scheme. 


%

In this paper, we propose a robust time-domain BA scheme for wideband mmWave channels. In our scheme, the mmWave channel sparsity in both angle and delay domains \cite{NasserAngleTime2016} is exploited to reduce the training overhead. Each user independently estimates its best AoA-AoD over the reserved beacon slots (see Section \ref{systemmodel}), during which the BS periodically broadcasts its \textit{Pseudo-Noise} (PN) sequences  (other orthogonal sequences already used in existing {\em Code Division Multiple Access} (CDMA) multiplexing technique, e.g., Gold sequences or Kasami sequences, are also feasible) according to a  beamforming codebook. Instead of estimating the instantaneous channel as done in \cite{AhmedFreqOMP2015,Gaozhen2016,LeeTwoStage2016,AhmedTime2017,AlkhateebTimeDomain2017}, we focus on estimating the channel second order statistics such that the estimation is more robust to time variations. We show that the resulting BS problem boils down to a {\em Non-Negative Least Squares} (NNLS) problem which can be efficiently solved by standard techniques. It is shown through simulation results that the proposed scheme is more efficient to estimate the best AoA-AoDs for each user and more robust to different time variations. Moreover, the proposed scheme is particularly beneficial for multi-user mmWave systems, where all the users within the BS coverage can be trained simultaneously.

{\em Notation:} In this paper, we use $\bfA^\transp$, $\bfA^*$, and $\bfA^\herm$ for transpose, conjugate, and conjugate transpose of the matrix $\bfA$. The Kronecker product of two matrices is denoted as $\bfA \otimes \bfB$. For an positive integer $k\in\intgr^+$, we use the shorthand notation $[k]$ for the set of non-negative integers $\{1,...,k\}$.

%
%
%

\section{System Model}\label{systemmodel}

\subsection{Channel Model}\label{channelmode}
\begin{figure}[t]
	\centerline{\includegraphics[width=8.5cm]{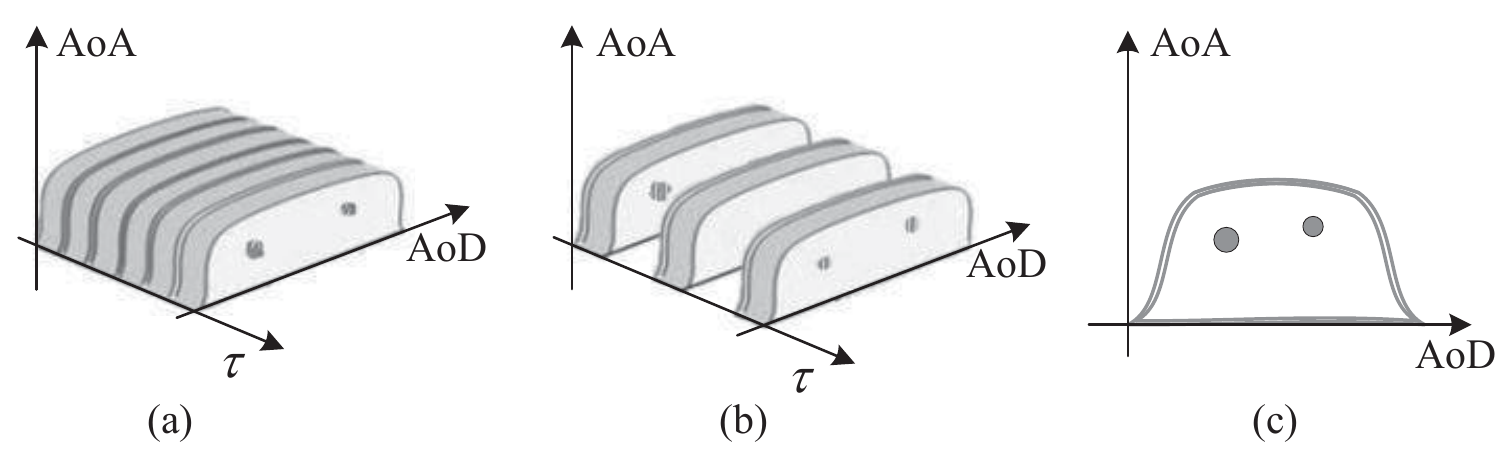}}
	\caption{{\small  Illustration of the channel sparsity in the {\em Angle of Arrival} (AoA), {\em Angle of Departure} (AoD) and delay ($\tau$) domains. (a) Power intensity in the joint AoA-AoD-Delay domain. (b) Power intensity over discrete delay taps. (c) Power intensity in AoA-AoD domain after an integration along all the delay taps.}}
	\label{1_bread}
\end{figure}

Consider a mmWave system with a {\em Base Station} (BS) and a generic user. The communication between the BS and the user occurs via a collection of sparse {\em Multi-Path Components} (MPCs) in the AoA-AoD-delay $(\theta,\phi,\tau)$ domain, as shown in \figref{1_bread} (a), where the dark spots inside the ``toast'' represent the strong scatterers connecting the BS and the user. Each scatterer can be regarded as a cluster consisting of multiple small components. In many frequency selective channels, the actual number of multipaths is limited. Consequently, the channel impulse response tends to be sparse in the time domain, and can be modeled as a tapped delay line (refer to \figref{1_bread} (b)) \cite{NasserAngleTime2016}. For BA, we just need the power intensity in the AoA-AoD domain, which can be obtained by taking an integration over all the delay taps, and finding the strongest scattering cluster in the AoA-AoD domain, as illustrated in \figref{1_bread} (c).

We assume that the BS has a {\em Uniform Linear Array} (ULA) with $M$ antennas and $M_{\text{RF}}\ll M$ RF chains. The user has a ULA with $N$ antennas and $N_{\text{RF}}\ll N$ RF chains. We assume that the scattering channel consists of  $L\ll \max\{M,N\}$ multi-path clusters each of which has a specific delay $\tau_l$ and a specific Doppler shift $\nu_l=\frac{\Delta v_{l}f_0}{c}$, where $\Delta v_{l}$ denotes the relative speed of the receiver, the $l$-th scatterer and transmitter, where $f_0$ and $c$ denote the carrier frequency and speed of light respectively \cite{OverviewHeath2016}. The $N\times M$ low-pass equivalent impulse response of the channel at slot $s$ is given by
\begin{align}\label{ch_mod_disc_mp}
\sfH_s(t,\tau)&=\sum_{l=1}^L \rho_{s,l} e^{j2\pi \nu_{l}t}\bfa_{\text{R}}(\phi_l) \bfa_{\text{T}}(\theta_l)^\herm \delta(\tau-\tau_l),
\end{align}
where $\rho_{s,l}$ is the random complex gain of the $l$-th cluster, where $(\phi_l, \theta_l, \tau_l)$ denote the (AoA, AoD, delay) of the $l$-th cluster, and where $\delta(\cdot)$ is the  Dirac delta function. The vectors $ \bfa_{\text{T}}(\theta_l)$  and $ \bfa_{\text{R}}(\phi_l)$ are the array response vectors of the BS at AoD $\theta_l$ and the user at AoA $\phi_l$  respectively.
%
%
We adopt a block fading model, where the channel gain $\rho_{s,l}$, $l \in [L]$, remains invariant over the channel \textit{coherence time} $\Delta t_c$ but changes randomly across different \textit{coherence times}. We assume that each MPC is a superposition of many smaller components that have (roughly) the same AoA-AoD and delay, such that the channel gain ${\rho_{s,l} \sim \cg(0, \gamma_l)}$ has a zero-mean complex Gaussian distribution.

Moreover, we are interested in the sparse channel matrix in the virtual angle domain \cite{sxsBA2017} which can be written as
\begin{align}\label{beamspacechannel}
\check{\sfH}_s(t,\tau) = \bfF_{N}^\herm\sfH_s(t,\tau)\bfF_{M}
\end{align}
where $\bfF_{N}\in\Co^{N\times N}$ and  $\bfF_{M}\in\Co^{M\times M}$ are the {\em Discrete Fourier Transformation} (DFT) matrices obtained by quantizing the virtual angle with the resolutions of $2\pi/N$ at the user and $2\pi/M$ at the BS, respectively. The sparse nature of the mmWave channel is explicitly reflected in the sparse nature of the angle domain channel matrix $\check{\sfH}_s(t,\tau)$, i.e., $\check{\sfH}_s(t,\tau)$ can be approximated as a $L$-sparse matrix, with $L$ non-zero elements in the positions corresponding to the AoA-AoDs of the $L$ scatterers. There is a grid error in \eqref{beamspacechannel}, since the AoAs/AoDs do not necessarily fall into the uniform grid. Nevertheless, this error becomes negligible by increasing the number of antennas \cite{sxsBA2017}.

\subsection{Singling Procedure}\label{signaling}
We assume that the BS can simultaneously transmit up to $M_{\text{RF}}\ll M$ different data streams. Similar to CDMA multiplexing technique, we assign a unique \textit{Pseudo-Noise} (PN) sequence to each RF chain. Let $x_{s,i}(t)$, $t \in [st_0, (s+1)t_0)$, be the continuous-time baseband equivalent PN signal corresponding to the $i$-th ($i\in[M_{\text{RF}}]$) data stream, where $t_0\leq \Delta t_c$ denotes the duration of the PN sequence. 
We assume that the PN sequence has a chip duration of $T_c$, a bandwidth of $B'=1/T_c \leq B$,  where $B$ is the maximum available bandwidth, and has a total of $N_c=t_0/T_c=t_0B'$ chips.

%
%

For each RF chain, the BS applies a normalized beamforming vector $\bfu_{s,i} \in \bC^M$ to $x_{s,i}(t)$. The transmittd signal at slot $s$ is given by
\begin{align}
\bfx_s(t) & =\sum_{i=1}^{M_{\text{RF}}} x_{s,i}(t) \bfu_{s,i}.
\end{align}
The received baseband equivalent signal at the user array is
\begin{align}\label{receiveTT}
\bfr_s(t)&=\int \sfH_s(t,d\tau) \bfx_s(t-\tau) \nonumber \\
&=\sum_{l=1}^L \rho_{s,l} e^{j2\pi \nu_{l}t} \bfa_{\text{R}}(\phi_l) \bfa_{\text{T}}(\theta_l)^\herm \bfx_s(t-\tau_l) \nonumber \\
&:=\sum_{l=1}^L \sfH_{s,l}(t)\bfx_s(t-\tau_l)\nonumber\\
&=\sum_{i=1}^{M_{\text{RF}}} \sum_{l=1}^L  \sfH_{s,l}(t)\bfu_{s,i} x_{s,i}(t-\tau_l),
\end{align}
where $\sfH_{s,l}(t) = \rho_{s,l} e^{j2\pi \nu_{l}t} \bfa_{\text{R}}(\phi_l) \bfa_{\text{T}}(\theta_l)^\herm$, $l\in[L]$ are the time-varying channels (due to the Doppler shift) corresponding to the $L$ independent scattering clusters.

This analog signal at the user side is split into $N_{\text{RF}}$ chains, which divides the signal power by a factor of $N_{\text{RF}}$. It follows that the received signal at the output of the $j$-th RF chain, $j\in[N_{\text{RF}}]$, is given by
\begin{align}\label{eq:j_out}
y_{s,j}(t)&=\frac{1}{\sqrt{N_{\text{RF}}}} \bfv_{s,j}^\herm \bfr_s(t)+z_{s,j}(t)\nonumber\\
&=\sum_{i=1}^{M_{\text{RF}}}\!\frac{\bfv_{s,j}^\herm}{\sqrt{N_{\text{RF}}}}\!\sum_{l=1}^L\sfH_{s,l}(t)x_{s,i}(t\!-\!\tau_l)\bfu_{s,i}+z_{s,j}(t),
\end{align}
where ${\bfv_{s,j} \in \bC^N}$ denotes the normalized beamforming vector of the $j$-th RF chain at the user side, $z_{s,j}(t)$ is the continuous-time complex \textit{Additive White Gaussian Noise} (AWGN) at the output of the $j$-th RF chain, with a \textit{Power Spectral Density} (PSD) of $N_0$ Watt/Hz.

\begin{figure}[t]
	\centerline{\includegraphics[width=9cm]{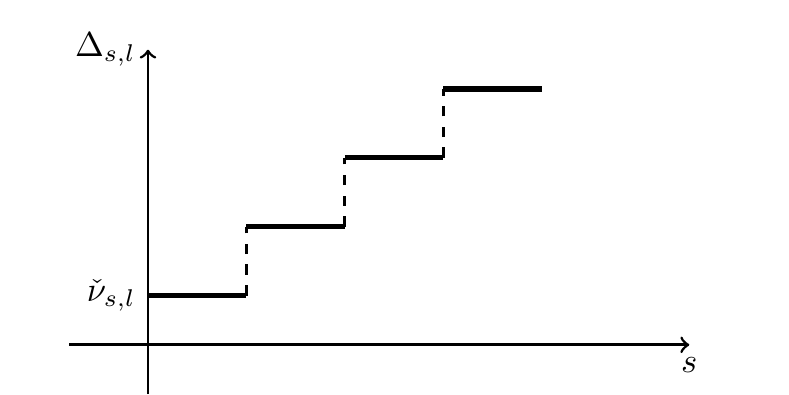}}
		\caption{{\small Illustration of the piecewise constant Doppler shift $\Delta_{s,l}=\check{\nu}_{s,l}+\nu_{l}t_0s$ with an initial $\check{\nu}_{s,l}$ and a constant step height $\nu_{l}t_0$, where $\check{\nu}_{s,l}$ changes i.i.d randomly over different beacon slots.}}
	\label{DopplerShift}
\end{figure}

We assume that each RF chain at the BS side is assigned a unique PN sequence, such that the $M_{\text{RF}}$ data streams transmitted from the BS side can be approximately separated at the user by correlating the received signal with a desired matched filter $x_{i}^*(-t)$. We obtain
the $i$-th BS data stream at the $j$-th RF chain of the user with receive beamforming vector $\vv_{s,j}$ by
\begin{align}\label{eq:j_outFrom_i}
y_{s,i,j}(t)&=\int y_{s,j}(\tau)x_{s,i}^*(\tau-t)d\tau\nonumber\\
&\overset{(a)}{\approx}\frac{\bfv_{s,j}^\herm}{\sqrt{N_{\text{RF}}}}\!\sum_{l=1}^L\sfH_{s,l}R_{i,i}^x(t\!-\!\tau_l)\bfu_{s,i}+z_{s,j}^c(t)
\end{align}
where $R_{i,i}^x(t)=\int x_{s,i}(\tau)x_{i}^*(\tau-t)d\tau$ is the autocorrelation of $x_{i}(t)$ at slot $s$, where  $z_{s,j}^c(t) = \int z_{s,j}(\tau)x_{i}^*(\tau-t)d\tau$  denotes the noise at the output of the matched filter, and where $(a)$ follows the fact that the cross-correlations between different PN sequences are nearly zero, i.e., $R_{i'\!,i}^x(t)=\int x_{s,i'}(\tau)x_{i}^*(\tau-t)d\tau\approx 0$ if $i'\neq i$. Here we used the approximation that $\sfH_{s,l}(t)\approx\rho_{s,l} e^{j2\pi (\check{\nu}_{s,l}+\nu_{l}t_0s)} \bfa_{\text{R}}(\phi_l) \bfa_{\text{T}}(\theta_l)^\herm$ is almost constant over the whole PN sequence,  where $\check{\nu}_{s,l}$ changes i.i.d randomly over different beacon slots, and where the overall shift $\Delta_{s,l}=\check{\nu}_{s,l}+\nu_{l}t_0s$ follows a piecewise constant procedure as shown in \figref{DopplerShift}. Hence for simplicity, we ignore the index $t$ in $\sfH_{s,l}(t)$.

We assume that the signal is sampled at chip-rate, where the discrete-time version of \eqref{eq:j_outFrom_i} is given by
\begin{align}\label{eq:j_outFrom_i_sample}
y_{s,i,j}[k]&=y_{s,i,j}(t)|_{t=kT_c}\nonumber\\
&=\frac{\bfv_{s,j}^\herm}{\sqrt{N_{\text{RF}}}}\!\sum_{l=1}^L\sfH_{s,l}R_{i,i}^x(kT_c\!-\!\tau_l)\bfu_{s,i}+z_{s,j}^c[k],
\end{align}
where we assume $k\in[\check{N}_c]$, $\check{N}_c\geq N_c+\Delta \tau_{\max}$, and where $\Delta \tau_{\max}=\max_{l,l'\in[L]}\{|\tau_l-\tau_{l'}|/T_c\}$. We refer to \figref{1_bread} (b), as we can see that, $y_{s,i,j}[k]$ corresponds to the projection of the $k$-th slice of  power intensity in the delay domain.

\section{Proposed Scheme}

\begin{figure}[t]
	\centerline{\includegraphics[width=8.0cm]{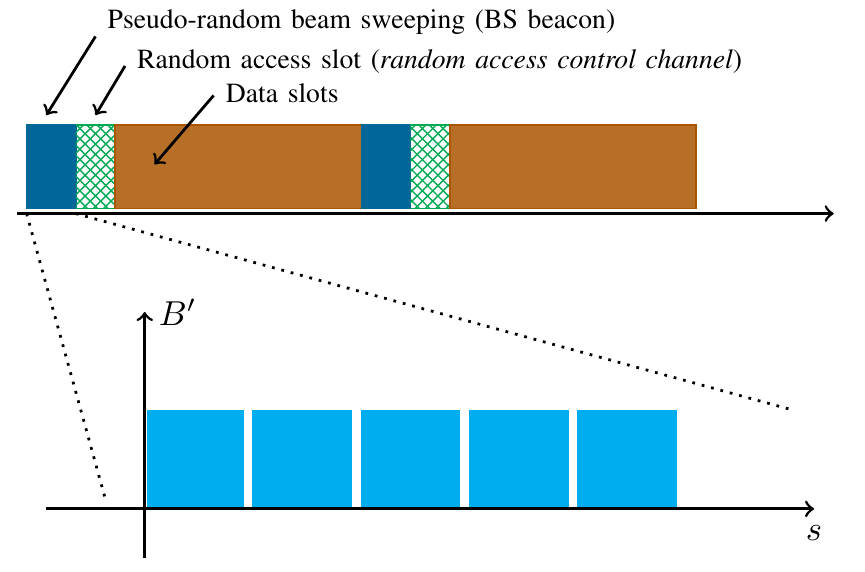}}
		\caption{{\small (Top) Frame structure of the proposed \textit{Beam Alignment} (BA) scheme at the BS side. (Bottom) Each beacon slot indexed by $s$ consists of $S$ PN sequences, and all the PN sequences have access to the whole effective bandwidth $B'\leq B$. Different data streams are distinguished by different PN sequences.}}
	\label{2_framestructure}
\end{figure}

\figref{2_framestructure} shows the frame structure of our proposed scheme. During the downlink {\em Beacon Slots}, the BS periodically broadcasts a probing signal that consists of $M_{\text{RF}}$ PN sequences to all the users, while all the users stay in the listening mode. The probing signal is transmitted along some random discrete angles determined by a {\em pseudo-random beamforming codebook}, which is priori known to all users. When a user, depending on its own channel condition, has gathered enough measurements, it estimates the AoA-AoD of its strongest path, and then feeds this information back to the BS over the uplink {\em random access slot}, during which the BS stays in a sectorized listening mode. We assume that, with high probability, the beamforming gain at the user and the sectored gain at the BS are sufficiently large to successfully decode the feedback information. More details are provided in the following sections.  

\subsection{Sparse Formulation}
As explained in Section \ref{channelmode} \eqref{beamspacechannel}, it is convenient to write \eqref{eq:j_outFrom_i_sample} directly in terms of the angle domain representation as 
\begin{align}\label{eq:j_outFrom_i_sample_beamspace}
y_{s,i,j}[k]=\frac{\check{\bfv}_{s,j}^\herm}{\sqrt{N_{\text{RF}}}}\!\sum_{l=1}^L\check{\sfH}_{s,l}R_{i,i}^x(kT_c\!-\!\tau_l)\check{\bfu}_{s,i}+z_{s,j}^c[k],
\end{align}
where the variables with `` $\check{}$ '' are the projections via $\Fm_M$ and $\Fm_N$ correspondingly. More specifically, $\check{\sfH}_{s,l} = \bfF_{N}^\herm\sfH_{s,l}\bfF_{M}$. We define $\check{\bfu}_{s,i}$ as a $M\times1$ vector with $\kappa_u$ non-zero values at components that correspond to the probing directions indexed by the {\em pseudo-random beamforming codebook} and zeros elsewhere. For simplicity, the non-zero values are equally set as $1/\sqrt{\kappa_u}$. Hence the $i$-th beamforming vector at the BS side is given by $\uv_{s,i} = \Fm_M \check{\uv}_{s,i}$. Similarly, at the user side, we define $\check{\bfv}_{s,j}$ as a $N\times1$ vector with $\kappa_v$ non-zero values $1/\sqrt{\kappa_v}$  at the components that randomly selected in $\check{\vv}_{s,j}$. The resulting beamforming vector at the $j$-th RF chain is given by $\vv_{s,j} = \Fm_N \check{\vv}_{s,j}$. 


To formulate the CS problem, we define $\check{\bfh}_{s,l}\! =\! 1/\sqrt{N_{\text{RF}}}\cdot\vec{(\check{\sfH}_{s,l})}$, $\check{\bfH}_s=[\check{\bfh}_{s,1}\,\cdots\, \check{\bfh}_{s,L}]$, $\bfc^{i}_k=[R_{i,i}^x(kT_c\!-\!\tau_1)\,\cdots\,R_{i,i}^x(kT_c\!-\!\tau_L)]^\transp$. We can then express \eqref{eq:j_outFrom_i_sample_beamspace} as
\begin{align}\label{eq:j_out_beamspace}
y_{s,i,j}[k]&=\frac{\check{\bfv}_{s,j}^\herm}{\sqrt{N_{\text{RF}}}}\!\sum_{l=1}^L\check{\sfH}_{s,l}R_{i,i}^x(kT_c\!-\!\tau_l)\check{\bfu}_{s,i}+z_{s,j}^c[k]\nonumber\\
&=(\check{\bfu}_{s,i}\otimes \check{\bfv}^*_{s,j})^\transp \check{\bfH}_s \bfc^{i}_k+z_{s,j}^c[k]\nonumber\\
&:=\bfg_{s,i,j}^\transp \check{\bfH}_s \bfc^{i}_k+z_{s,j}^c[k].
\end{align}
We assume the channel gains $\rho_{s,l}$ in $\check{\bfH}_s$ change i.i.d randomly across different beacon slots. Note that for the high symbol rates at mmWaves (for example, the chip rate used in IEEE 802.11ad preamble is 1760MHz) \cite{AhmedTime2017}, it is impractical to use different beamforming vectors for different symbols within the same beacon slot. We assume that each beacon slot contains $S$ symbols, during which the combined probing vector $\bfg_{s,i,j}$ remains constant whereas $\check{\bfH}_s$ only changes because of the Doppler shifts $\nu_{l}$ (refer to \figref{DopplerShift}). We assume the probing vector $\bfg_{s,i,j}$ changes across different slots according to our {\em pseudo-random beamforming codebook}. We index the received symbols belonging to the $(s+1)$-th beacon slot as $sS+s'$, $s'\in[S]$. It follows that the received signal through the $i$-th RF chain at the BS and the $j$-th RF chain at the user after matched filtering (refer to \eqref{eq:j_out_beamspace}) can be written as
\begin{align}\label{eq:j_out_beamspaceT1T2}
y_{sS+s',i,j}[k]=\bfg_{s,i,j}^\transp \check{\bfH}_{sS+s'} \bfc^{i}_k+z_{sS+s',j}^c[k].
\end{align}

To ensure our estimation is more robust to different time variations, we focus on the second order statistics of the channel coefficients. More specifically, we compute the energy at the output of the matched filter as
\begin{align}\label{q_check}
&\check{q}_{sS+s',i,j}=\sum_{k=1}^{\check{N}_c}|y_{sS+s',i,j}[k]|^2\nonumber\\
&=\bfg_{s,i,j}^\transp\left(\sum_{l=1}^{L}\check{\bfh}_{sS+s',l}\check{\bfh}_{sS+s',l}^\herm\sum_{k=1}^{\check{N}_c}|R_{i,i}^x(kT_c\!-\!\tau_l)|^2\right)\bfg_{s,i,j}\nonumber\\
&+\sum_{k=1}^{\check{N}_c}|z_{sS+s',j}^c[k]|^2+\sum_{k=1}^{\check{N}_c}\xi_{sS+s',i,j}^{h}+\sum_{k=1}^{\check{N}_c}2\xi_{sS+s',i,j}^{z},
\end{align}
where $\check{N}_c\geq N_c+\Delta \tau_{\max}$, and where the first and second terms represent signal and noise contributions respectively. Note that in the signal part, only $L$ out of $\check{N}_c$ slices which correspond to the $L$ scatterers in the delay domain contains signal power, while all the other slices are approximately zero due to the low cross-correlation property of PN sequences.  Moreover, the last two parts in \eqref{q_check} denoting the cross-terms are given by
\begin{align}
\xi_{s,i,j}^{h} &=\sum_{l\neq l'}^L \bfg_{s,i,j}^\transp\check{\bfh}_{sS+s',l}\check{\bfh}_{sS+s',l'}^\herm R_{i,i}^x(kT_c\!-\!\tau_l)\nonumber\\
&\times R_{i,i}^x(kT_c\!-\!\tau_{l'})^\herm\bfg_{s,i,j},
\end{align}
\begin{align}
\xi_{sS+s',i,j}^z=2\Re\left \{ \bfg_{s,i,j}^\herm \check{\bfH}_{sS+s'} \bfc^{i}_k \cdot z_{sS+s',j}^c[k]^\herm \right \}.
\end{align}

To obtain a more reliable statistic measurement, we take an average of \eqref{q_check} over the $S$ sequences within one beacon slot. The key idea underlying our method follows the fact that, when the energy averaging dimensions $S$ is sufficiently large, the cross terms are negligible. Hence we have
\begin{align}\label{q_nocheck}
&q_{s,i,j}=\frac{1}{S}\sum_{s'=1}^{S} \check{q}_{sS+s',i,j}\nonumber\\
&\approx \frac{\bfg_{s,i,j}^\transp}{S}\!\sum_{s'=1}^{S}\!\!\left(\!\sum_{l=1}^{L}\check{\bfh}_{sS+s'\!,l}\check{\bfh}_{sS+s'\!,l}^\herm\!\sum_{k=1}^{\check{N}_c}|R_{i,i}^x(kT_c\!-\!\tau_l)|^2\!\right)\bfg_{s,i,j}\nonumber\\
&+\frac{1}{S}\sum_{s'=1}^{S}\left(\sum_{k=1}^{\check{N}_c}|z_{sS+s',j}^c[k]|^2\right).
\end{align}

Without loss of generality, we assume the energy contained in each PN sequence is constant denoted by $R^x(0)=R_{i,i}^x(0)=\frac{\ptot t_0}{M_{\text{RF}}}=\frac{\ptot N_cT_c}{M_{\text{RF}}}=\frac{\ptot N_c}{M_{\text{RF}}B'}$, $\forall i\in[M_{\text{RF}}]$, where $\ptot$ is the total transmit power from the BS, where $t_0$ is the PN sequence interval with chip duration $T_c$ and a total of $N_c$ chips, and where the effective bandwidth $B'=1/T_c$. Defining a $N\times M$ matrix $\Gammam$ that coincides with the second order statistics of the channel coefficients, with 
\begin{align}
[\Gammam]_{n,m} = \sum_{l=1}^L\bE\left[|[\check{\sfH}_{sS+s',l}]_{n,m}|^2\right]\cdot \frac{|R^x(0)|^2}{\kappa_u\kappa_v N_{\text{RF}}}.
\end{align} 
Also, from $\frac{1}{S}\sum_{s'=1}^{S}|z_{sS+s',j}^c[k]|^2\to\bE[|z_{sS+s',j}^c[k]|^2]=N_0R^x(0)$, we can approximate \eqref{q_nocheck} by
\begin{align}\label{q_nocheck3}
q_{s,i,j} = \bfb_{s,i,j}^\transp \vec{(\Gammam)} + \check{N}_cN_0R^x(0) + w_{s,i,j},
\end{align}
where $\bfb_{s,i,j} = \bfg_{s,i,j}\sqrt{\kappa_u\kappa_v}$ denotes the binary combined probing window (a $MN\times 1$ vector), containing $1$ at the probed AoA-AoD components and $0$ elsewhere. We consider a residual term  $w_{s,i,j}$, such that 
the approximation in \eqref{q_nocheck} yields the equality in \eqref{q_nocheck3}.
%

Following this procedure, over $T$ beacon slots the user obtains a total number of $M_{\text{RF}}N_{\text{RF}}T$  equations, which can be written in the form
\begin{align}\label{UE_equations}
\bfq=\bfB \cdot\vec{(\Gammam)} + \check{N}_cN_0R^x(0)\cdot \one + \bfw,
\end{align}
where the vector ${\bfq=[q_{1,1,1}, \dots q_{1,M_{\text{RF}},N_{\text{RF}}}, \dots, q_{T,M_{\text{RF}},N_{\text{RF}}}]^\transp}$ consists of all $M_{\text{RF}}N_{\text{RF}}T$ measurements achieved as in (\ref{q_nocheck3}), ${\bfB=[\bfb_{1,1,1}, \dots, \bfb_{1,M_{\text{RF}},N_{\text{RF}}}, \dots, \bfb_{T,M_{\text{RF}},N_{\text{RF}}}]^\transp}$ is uniquely defined by the {\em pseudo-random beamforming codebook} of the BS and the local beamforming codebook of the user, and  $\bfw \in \bR^{M_{\text{RF}}N_{\text{RF}}T}$ denotes the residual fluctuations. 

For later use, we define the SNR per chip during training phase in \eqref{eq:j_out_beamspaceT1T2} as
\begin{align}\label{snrPerchip}
\snr^{y_{s,i,j}[k]} =& \frac{|R^x(0)|^2\sum_{l=1}^{L}\gamma_l\cdot \one_{\{kt_p=\tau_l\}}}{\bE[|z_{s,j}^c[k]|^2]\cdot \kappa_u\kappa_vN_{\text{RF}}}\nonumber\\
& = \frac{|R^x(0)|^2\sum_{l=1}^{L}\gamma_l\cdot \one_{\{kt_p=\tau_l\}}}{N_0R^x(0)\cdot \kappa_u\kappa_vN_{\text{RF}}}\nonumber\\
&=\frac{R^x(0)\sum_{l=1}^{L}\gamma_l\cdot \one_{\{kt_p=\tau_l\}}}{\kappa_u\kappa_vN_{\text{RF}}N_0}\nonumber\\
&=\frac{\ptot t_0\sum_{l=1}^{L}\gamma_l\cdot \one_{\{kt_p=\tau_l\}}}{\kappa_u\kappa_vM_{\text{RF}}N_{\text{RF}}N_0}\nonumber\\
&=\frac{\ptot N_cT_c\sum_{l=1}^{L}\gamma_l\cdot \one_{\{kt_p=\tau_l\}}}{\kappa_u\kappa_vM_{\text{RF}}N_{\text{RF}}N_0}\nonumber\\
&=\frac{\ptot N_c\sum_{l=1}^{L}\gamma_l\cdot \one_{\{kt_p=\tau_l\}}}{\kappa_u\kappa_vM_{\text{RF}}N_{\text{RF}}N_0B'},
\end{align}
where $\one_{\{kt_p=\tau_l\}}$ is the indicator function. Note that taking $t_0=N_cT_c=N_c\cdot \frac{1}{B'}\leq \Delta t_c$ as a prerequisite, if the PN sequence length $N_c$ is fixed, larger chip duration $T_c$ (i.e., larger $t_0$) implies an increase of the SNR in \eqref{eq:j_out_beamspaceT1T2}\eqref{snrPerchip} but at the cost of less effective bandwidth $B'=1/T_c\leq B$.   

We define the SNR in \eqref{q_nocheck} as
\begin{align}\label{snrSum}
\snr^{q_{s,i,j}} &=\frac{\ptot N_c\sum_{l=1}^{L}\gamma_l}{\check{N}_c\kappa_u\kappa_vM_{\text{RF}}N_{\text{RF}}N_0B'}\nonumber\\
&\overset{(a)}{\approx} \frac{\ptot\sum_{l=1}^{L}\gamma_l}{\kappa_u\kappa_vM_{\text{RF}}N_{\text{RF}}N_0B'}
\end{align}
where $(a)$ follows $\check{N}_c\geq N_c+\Delta \tau_{\max}$, the relative difference between $\check{N}_c$ and $N_c$ is very small when $N_c$ is large.

We also define the SNR before beamforming by
\begin{align}\label{snrBBF}
\snrbef =\frac{\ptot \sum_{l=1}^{L}\gamma_l}{MNN_0B}.
\end{align}
This is the SNR obtained when a single data stream ($M_{\text{RF}} = 1$)  is transmitted through a single BS antenna and 
is received in a single user antenna  (isotropic transmission) over a single RF chain ($N_{\text{RF}} = 1$) with full-band spreading $B$.

\subsection{Channel Estimating via Non-Negative Least Squares}
At the user side, each user needs to recover the second order channel coefficients $\Gammam$ in \eqref{UE_equations} to find the strongest component. The index of the strongest component in $\Gammam$ implies the best AoA-AoD combination. As we have explained in our previous paper \cite{sxsBA2017}, recent progress in CS shows that when the underlying parameter $\Gammam$ is non-negative as in our case, the simple \textit{Least Squares} (LS) given by
\begin{align}\label{eq:NNLS}
\Gammam^*=\argmin_{\Gammam \in \bR_+^{N\times M}} \|\bfB \cdot\vec{(\Gammam)} + \check{N}_cN_0R^x(0)\cdot \one - \bfq\|^2, 
\end{align}
is sufficient to ensure a suitable recovery of $\Gammam$. The (convex) optimization problem \eqref{eq:NNLS} is generally referred to as \textit{Non-Negative Least Squares} (NNLS) and has been well investigated in the literature. The estimation dimension
of this NNLS problem only depends on the quantized angle
grid. Existing techniques such as Gradient Projection, Primal-Dual, etc., can efficiently solve this problem. We refer to \cite{nguyen2015anti}  for the recent progress on the numerical solution of 
NNLS and a discussion on other  related work in the literature.

\section{Numerical Results}
In this section, we evaluate the performance of our scheme via numerical simulations. We also compare our scheme with another recently proposed time-domain BA approach \cite{AlkhateebTimeDomain2017,AhmedTime2017} which focuses on estimating the instantaneous channel coefficients based on {\em Orthogonal Matching Pursuit} (OMP) technique. 

We consider a system with $M=32$ antennas, $M_{\text{RF}}=3$ RF chains at the BS side, and $N=32$ antennas, $N_{\text{RF}}=3$ RF chains at the user side. We assume a short preamble structure used in IEEE 802.11ad \cite{AhmedTime2017,ParameterPerahia2010}, where the beacon slot is of duration $t_0S=1.891\,\mu$s. The system is assumed to work at $f_0=70$ GHz, has a maximum available bandwidth of $B=1.76$ GHz as in \cite{AhmedTime2017,AlkhateebTimeDomain2017}. The relative speed $\Delta v_{l}$ for each path is around $1\sim5$ m$/$s. We announce an individual experiment to be successful if the index of strongest component in $\Gammam$, i.e., the actual scatterer location, is correctly estimated. In the following simulations, we would use \texttt{lsqnonneg.m} in\ \matlab  to solve the NNLS optimization \eqref{eq:NNLS}.

\subsection{Performance with different number of paths (scatterers) $L$}
\begin{figure}[t]
	\centerline{\includegraphics[width=8cm]{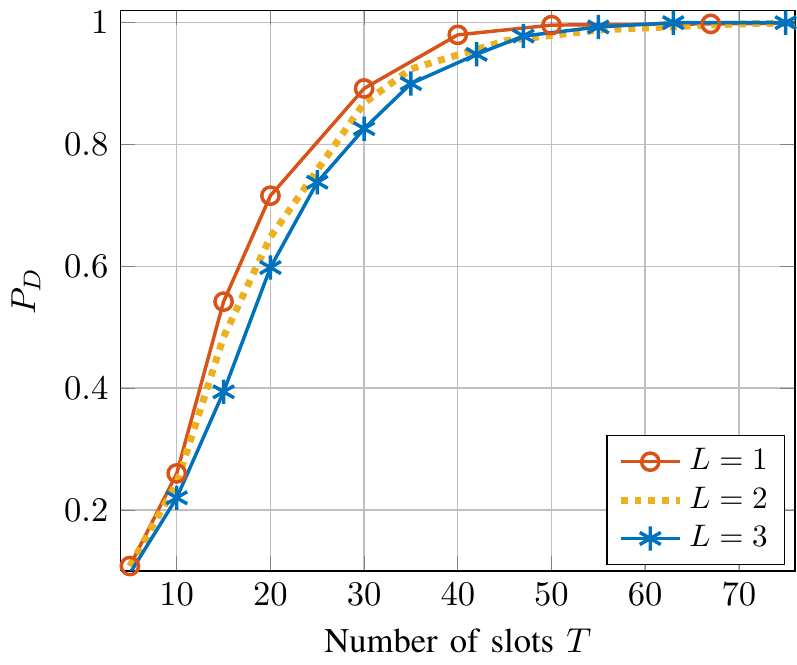}}
\caption{{\small  Detection probability $P_D$ of the proposed scheme for different number of paths (scatterers) $L$, where $M=N=32$, $M_{\text{RF}}=3$, $N_{\text{RF}}=2$, $\kappa_u=\kappa_v=16$, $B'=B$, $N_c=512$, $\snrbef=-15$ dB}.}
\label{changeScatter}
\end{figure}
To evaluate that the proposed scheme works equally efficiently for single-path ($L=1$) and multi-path ($L>1$) scenarios, we assume a set of parameters in our scheme, where the power spreading factor  $\kappa_u=\kappa_v=16$, effective bandwidth $B'=B$, PN sequence length $N_c=512$, SNR before beamforming $\snrbef=-15$ dB. \figref{changeScatter} shows the performance of our proposed scheme, where we announce an individual experiment to be successful if the strongest path is correctly estimated. As we can see from \figref{changeScatter},  the proposed scheme achieves similar performance for single path ($L=1$) and multi-path ($L=2,3$) scenarios, where in both cases, at most $T=50$ beacon slots (frames) is sufficient to ensure a successful beam alignment. In the following simulations, unless stated, we consider a single path $L=1$ scenario for the ease of illustration.

\subsection{Dependence on the beam spreading factors $(\kappa_u,\kappa_v)$}
\begin{figure}[t]
	\centerline{\includegraphics[width=8cm]{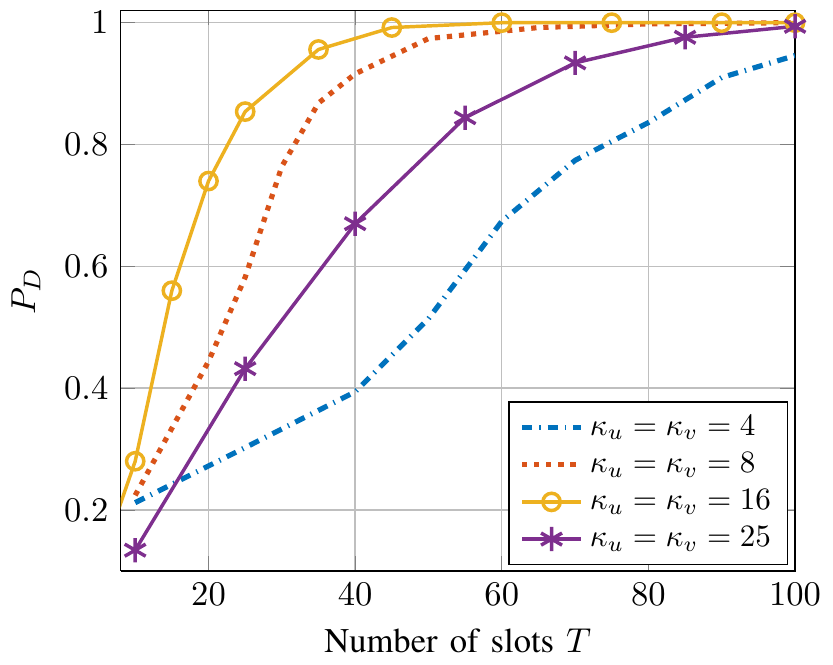}}
	\caption{{\small  Detection probability $P_D$ of the proposed scheme with respect to different power spreading factors ($\kappa_u$, $\kappa_v$), where $M=N=32$, $M_{\text{RF}}=3$, $N_{\text{RF}}=2$, $B'=B$, $N_c=512$, $S=6$, $\snrbef=-15$ dB}.}
	\label{changekvkuTD}
\end{figure}
\figref{changekvkuTD} shows the performance influences of the power spreading factors $\kappa_u$ and $\kappa_v$ in our proposed scheme. Since the power spreading factors $(\kappa_u,\kappa_v)$ impose a trade-off between the angle coverage of the probing window $\bfB$ and the SNR of the measurements \eqref{snrPerchip} \eqref{snrSum}, it can be seen from \figref{changekvkuTD} that, increasing the spreading factor from $\kappa_u=\kappa_v=4$ to $\kappa_u=\kappa_v=8,16$ improves the performance. However, the performance severely degrades when $(\kappa_u,\kappa_v)$ are increased to $\kappa_u=\kappa_v=25$.

\subsection{Dependence on the PN sequence length $N_c$}
\begin{figure}[t]
	\centerline{\includegraphics[width=8cm]{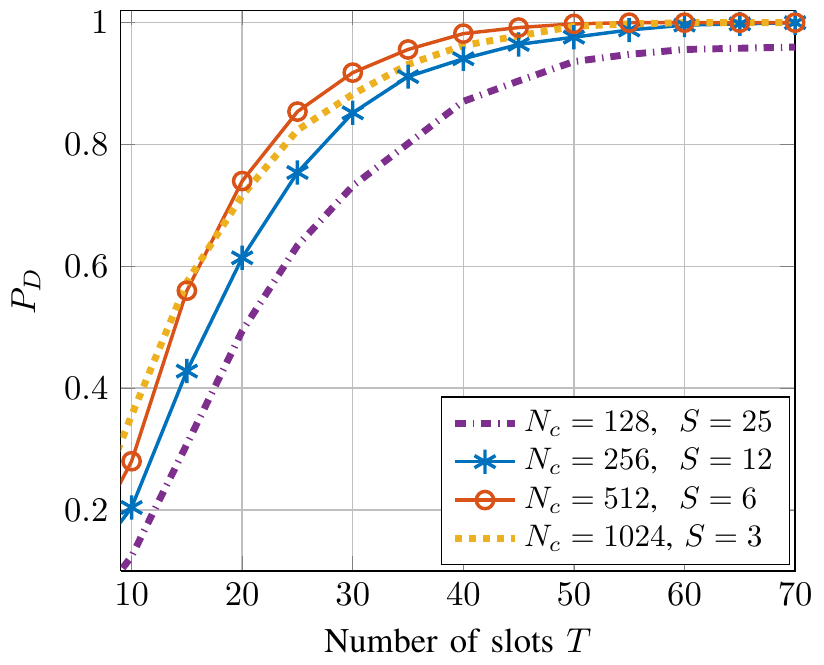}}
\caption{{\small  Detection probability $P_D$ of the proposed scheme for different PN sequence lengths $N_c$, where $M=N=32$, $M_{\text{RF}}=3$, $N_{\text{RF}}=2$, $\kappa_u=\kappa_v=16$, $B'=B$, $\snrbef=-15$ dB}.}
\label{changePnLengTD}
\end{figure}
For a wideband case, i.e., $B'=B$, chip duration $T_c=1/B$, each data stream has access to the whole bandwidth. When the duration of the beacon slot is fixed, which in our case is $N_c T_c S = 1.891\,\mu$s, \figref{changePnLengTD} shows that, increasing the PN sequence length $N_c$ from $N_c=128$ to $N_c=256,512$ improves the performance of our proposed scheme. However, the performance degrades slightly when $N_c$ is increased to $N_c=1024$. This is because the beacon slot is fixed, increasing $N_c$ implies decreasing $S$, which means less reliable statistic measurements as shown in \eqref{q_nocheck}.

\subsection{Comparison with other time-domain methods}
\begin{figure}[t]
	\centerline{\includegraphics[width=8cm]{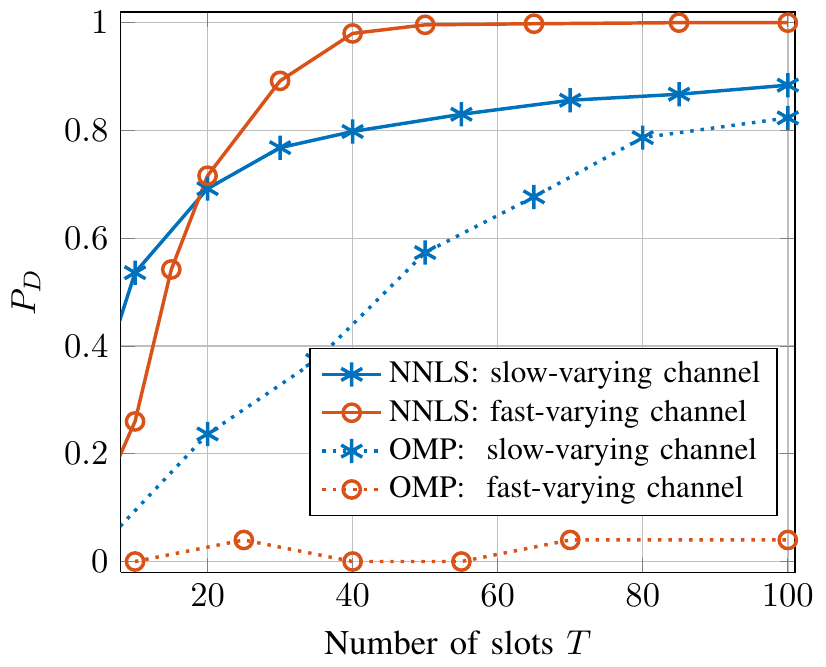}}
\caption{{\small Comparison of the proposed scheme based on NNLS with that in \cite{AlkhateebTimeDomain2017,AhmedTime2017} based on OMP  for both slow-varying and fast-varying channels, where $M=N=32$, $M_{\text{RF}}=3$, $N_{\text{RF}}=2$, $\kappa_u=\kappa_v=16$, $B'=B$, $N_c=512$, $\snrbef=-15$ dB}.}
\label{changeRHO}
\end{figure}
\figref{changeRHO} compares the performance of our proposed scheme and another recently proposed time-domain approach \cite{AlkhateebTimeDomain2017,AhmedTime2017} which is based on OMP technique. The approach in \cite{AlkhateebTimeDomain2017,AhmedTime2017} assumes that the instaneous channel remains constant and has no Doppler shifts (slow-varying channel) for the whole training stage. On the contrast, we focus on the channel second order statistics. We incorporate the random gains and Doppler shifts (fast-varying channel) into the channel matrix, which is more in line with the realistic mmWave channels. It can be seen from \figref{changeRHO} that, the proposed scheme exhibits much more robust performance whereas the approach in \cite{AlkhateebTimeDomain2017,AhmedTime2017} fails when the channel is fast time-varying.

\subsection{Dependence on the chip duration $T_c$}
\begin{figure}[t]
	\centerline{\includegraphics[width=8cm]{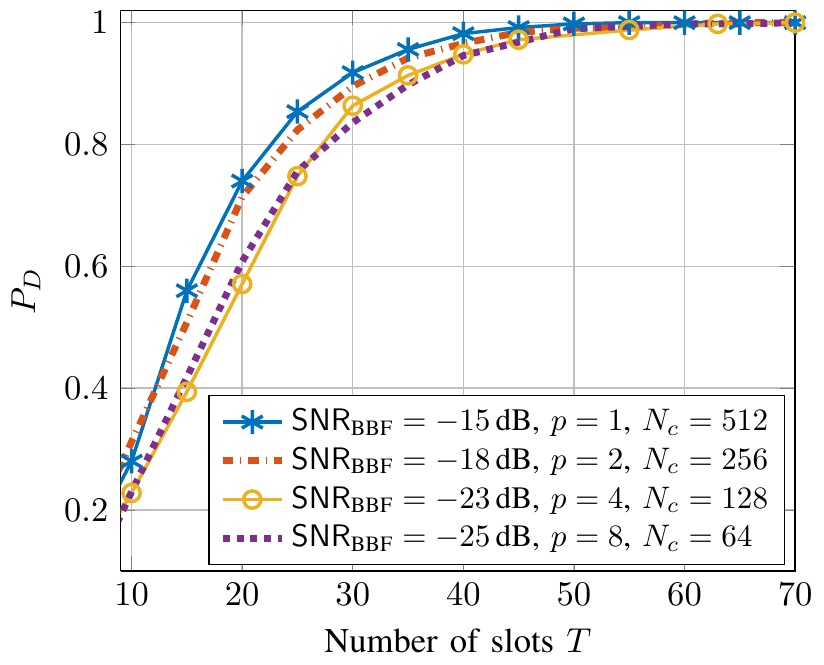}}
\caption{{\small  Detection probability $P_D$ of the proposed scheme for different $\snrbef$, where $M=N=32$, $M_{\text{RF}}=3$, $N_{\text{RF}}=2$, $\kappa_u=\kappa_v=16$, $B'=B/p$. Decreasing the SNR before beamforming $\snrbef$, similar performance is achieved by increasing the chip duration $T_c=1/B'=p/B$}.}
\label{changeChipDura}
\end{figure}
Note that another important parameter in our scheme is the chip duration $T_c=1/B' =p/B$, where $B$ is the maximum available bandwidth, and where we define $B'=B/p$ for the ease of illustration. When the duration of the PN sequence $t_0=N_cT_c\leq \Delta t_c$ is fixed, increasing the chip duration implies increasing $p$ (decreasing $B'$) and deceasing $N_c$. In that case, the measurement SNR in \eqref{q_nocheck} would be significantly improved (see, e.g., \eqref{snrSum}). Hence, even the SNR before beamforming \eqref{snrBBF} is very low, the proposed scheme can still achieve a good performance by increasing the chip duration $T_c$ as illustrated in \figref{changeChipDura}. Hence the fundamental parameter that really influences the performance is the code (PN sequence) interval $t_0$, i.e., once $t_0$ is fixed, we can use shorter code length $N_c$ with a longer chip duration $T_c$ as long as the effective bandwidth $B'=1/T_c$ is large enough such that we can resolve the delay elements.

%
%
\subsection{Power Delay Profile (PDP) before and after BA}
\begin{figure}[t]
	\centerline{\includegraphics[width=8.5cm]{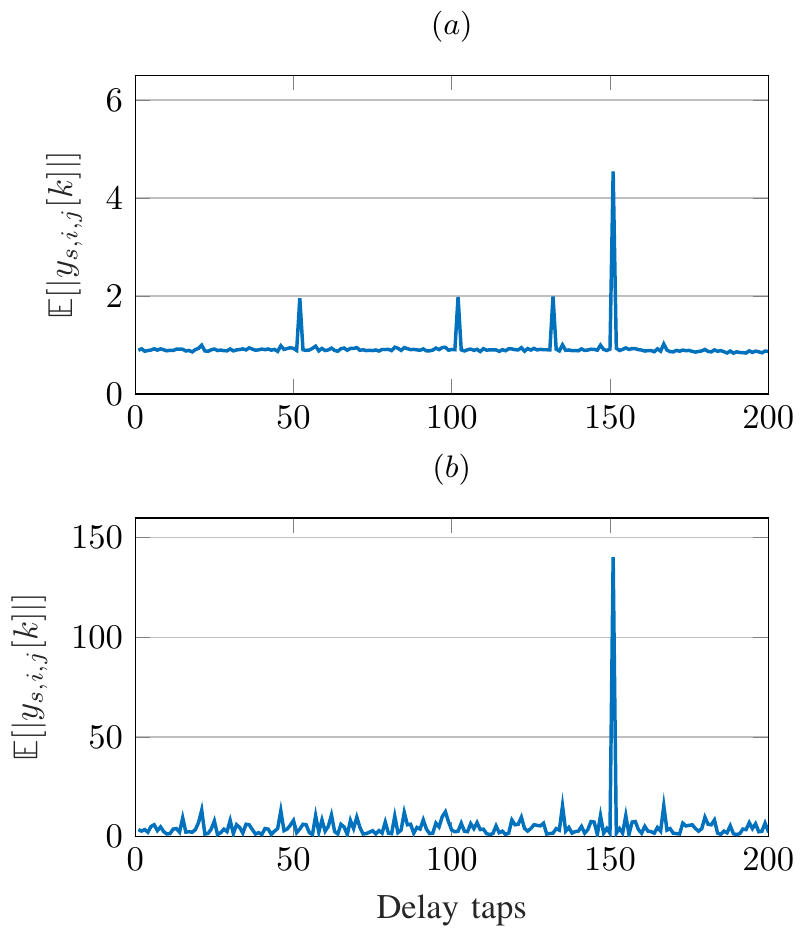}}
\caption{{\small Illustration of the {\em Power Delay Profile} (PDP) with multi-path ($L=4$) channel in \eqref{eq:j_outFrom_i_sample}. $(a)$ Before {\em Beam Alignment}. $(b)$ After {\em Beam Alignment}}}
\label{DIP}
\end{figure}
\figref{DIP} compares the average {\em Power Delay Profile} (PDP) of a mmWave channel with $L=4$ multipath components before and after BA. It can be seen from \figref{DIP} (a) that, before BA, the channel has a large delay spread and is highly frequency selective. Moreover, since different multi-path components are mixed with each other and since each has its own delay, the time-domain channel is highly time-varying. In contrast, as seen from \figref{DIP} (b), after BA, the channel effectively consists of a single multi-path component, thus, it is quite flat in frequency. Also, note that in contrast with the former case where different muli-path components were mixed with different Doppler frequencies, in the latter case the Doppler frequency of the single multi-path component can be easily compensated by the {\em Phase Locked Loop} (PLL) at the receiver. 

%
%
%
%
%

\section{Conclusion}
In this paper, we proposed a new time-domain {\em Beam Alignment} (BA) scheme. The proposed scheme is particularly beneficial for wideband multi-user mmWave systems, where each user has access to the whole bandwidth, and where all the users within the BS coverage can be trained simultaneously. We focused on the channel second order statistics, incorporating both the random channel gains and random Doppler shifts into the channel matrix to further approach the realistic mmWave channels. We applied the recently developed {\em Non-Negative Least Squares} (NNLS) technique to efficiently find the strongest path for each user. Simulation results showed that the proposed scheme required very few training overhead, and achieved very good robustness to fast time-varying channels.

\balance
{\footnotesize
\bibliographystyle{IEEEtran}
\bibliography{references}
}

\end{document}